\documentclass[printer]{aa}
\usepackage{psfig}

\newcommand{\am}[2]{$#1'\,\hspace{-1.7mm}.\hspace{.0mm}#2$}
\newcommand{\HI}{\mbox{H\,{\sc i}}}
\newcommand{\IHI}{\mbox{$I_{HI}$}}
\newcommand{\Jykms}{\mbox{Jy~km~s$^{-1}$}}
\newcommand{\kms}{\mbox{km~s$^{-1}$}}

\newcommand{\MsunLBsun}{\mbox{${M}_{\odot}$/$L_{\odot,B}$}}
\newcommand{\VHI}{\mbox{${V}_{HI}$}}

\begin{document}


\title{Galaxy transmutations: The double ringed galaxy ESO~474-G26\thanks{Based 
on observations made at the Observat\'{o}rio do Pico dos
Dias (OPD), operated by the MCT/Laborat\'{o}rio Nacional de Astrof\'{i}
sica,Brazil}}

\author{V. Reshetnikov\inst{1,}\inst{2,}\inst{5}
\and F. Bournaud\inst{2}
\and F. Combes\inst{2}
\and M. Fa\'{u}ndez-Abans\inst{3}
\and M. de Oliveira-Abans\inst{3}
\and W. van Driel\inst{4}
\and S.E. Schneider\inst{6}}

\offprints{resh@astro.spbu.ru }

\institute{Astronomical Institute of St.~Petersburg State 
University, 198504 St.~Petersburg, Russia 
\and
Observatoire de Paris, LERMA, 61 Av. de l'Observatoire, 75014 Paris, France
\and
MCT/Laborat\'{o}rio Nacional de Astrof\'{i}sica, Caixa Postal 21,
CEP:37.504-364, Itajub\'{a}, MG, Brasil
\and 
Observatoire de Paris, Section de Meudon, GEPI, 5 pl. Jules Janssen,
92195 Meudon, France
\and
Isaac Newton Institute of Chile, St.~Petersburg Branch
\and
University of Massachusetts, Astronomy Program, 536 LGRC, Amherst, 
MA 01003, U.S.A. 
}

\date{Received  / Accepted   2004}

\titlerunning{ESO~474-G26}

\authorrunning{Reshetnikov et al.}

\abstract{
Surface photometry and a 21cm \HI\ line spectrum of the giant double-ringed galaxy ESO~474-G26 are presented. The morphology of this system is unique among the 30,000 galaxies with $B\leq$15\fm5. Two almost orthogonal optical rings with diameters of 60 and 40 kpc surround the central body (assuming $H_0=70$~km~s$^{-1}$~Mpc$^{-1}$). The outer one is an equatorial ring, while the inner ring lies in a nearly polar plane. The rings have blue optical colors typical of late-type spirals. Both appear to be rotating around the central galaxy, so that this system can be considered as a kinematically confirmed polar ring galaxy. Its observational characteristics are typical of galaxy merger remnants. Although the central object has a surface brightness distribution typical of elliptical galaxies, it has a higher surface brightness for its effective radius than ordinary ellipticals. Possible origins of this  galaxy are discussed and numerical simulations are presented that illustrate the formation of the two rings in the merging process of two spiral galaxies, in which the observed appearance of ESO~474-G26 appears to be a transient stage.
\keywords{ galaxies: individual: ESO~474-G26 -- galaxies:
photometry -- galaxies: formation -- galaxies: structure}
}

\maketitle
\section{Introduction}

Polar ring galaxies (PRGs) and related objects are peculiar systems that
help to understand the formation of galaxies in general, since they
represent extreme cases, providing  clues on formation scenarios.
The clearest cases of polar rings, named ''kinematically confirmed'' objects
in the Polar Ring Catalog (PRC) of Whitmore et al. (1990),
are stable objects frozen in a peculiar morphology, with matter
rotating in two nearly perpendicular planes.
Numerical simulations have shown that this kind of system could be
explained either by simple gas accretion
(Reshetnikov \& Sotnikova 1997, and Bournaud \& Combes 2003, hereafter BC03),
or by mergers of perpendicularly oriented disk galaxies (Bekki 1997 \& 1998; BC03). However, in the PRC most objects
do not show such a simple appearance with an edge-on central object and an edge-on polar ring, but rather more complex morphologies, in which they are frozen in a quasi-equilibrium state, 
avoiding complete relaxation towards a fully mixed symmetric system. These systems with complex morphologies, called 
polar ring related objects, provide a unique chance to study PRG formation scenarios in objects in 
which the initial components have not yet disappeared -- if they are the results of mergers, it may  
still be possible to recognize the progenitors in the unrelaxed remnants, and reconstruct their formation  
events. This is no longer possible in more relaxed merger remnants, that have become elliptical galaxies 
with only faint shells and ripples.

In this article we present the results of photometric observations
and an \HI\ line spectrum of ESO~474-G26 (PRC C-3 in Whitmore et al. 1990, and AM~0044-243 in Arp \& Madore 1987), a unique galaxy with two almost orthogonal
large-scale optical rings (Fig.~1). ESO~474-G26 is in the list of most
luminous galaxies (Cappi et al. 1998), and a strong source
of far-infrared (NED\footnote{NASA/IPAC Extragalactic Database}) and 
CO emission (Galletta et al. 1997). Its nuclear spectrum is of an 
intermediate type -- LINER/HII (Sekiguchi \& Wolstencroft 
1993). The PRC shows  optical emission-line rotation curves along three 
position angles.

Observations are detailed in Sect.~2, and results in Sect.~3.
A numerical simulation is presented in Sect.~4 to illustrate
a possible formation mechanism.

\begin{figure}
\centerline{\psfig{file=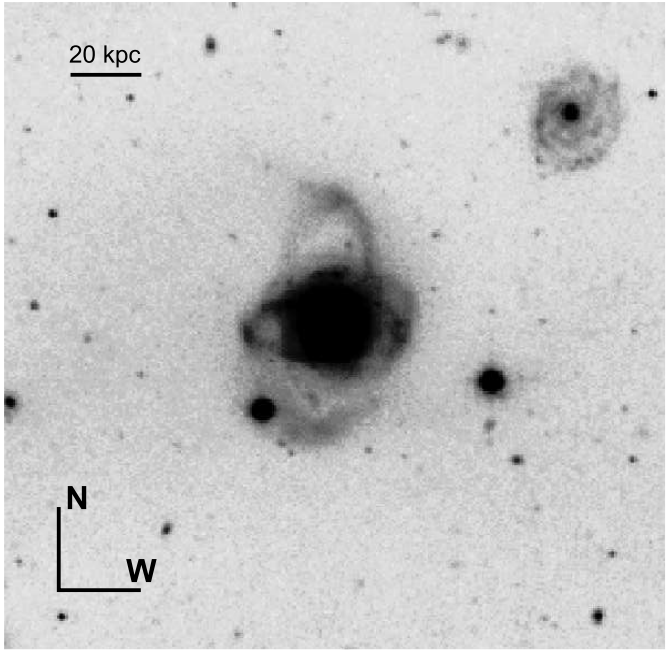,width=8.8cm,angle=0,clip=}}
\caption{$V$-band image of ESO~474-G26. The image size is  
$2\farcm8 \times 2\farcm8$. The spiral to the northwest is a background galaxy.}
\end{figure}

\section{Observations and data reduction}

\subsection{Photometric observations}

The photometric observations were obtained in August, 2002, on the 1.6-m
telescope of the Observat\'{o}rio do Pico dos Dias (operated by the MCT/Laborat\'{o}rio Nacional de Astrof\'{i}sica,Brazil), equipped with direct imaging 
camera \#1 and a thick, back-illuminated 2048x2048, 13.5-$\mu$m square 
pixels CCD detector \#98, and scale 0\farcs18/pixel.  
The readout noise was 2.4$e^{-}$ and the gain, 2.5$e^{-}$/ADU.

The data were acquired with standard Johnson $B$, $V$ and Cousins
$R$, $I$ band filters. Photometric calibration was made using standard stars from the Landolt (1983) and 
Graham (1982) lists. The seeing during the observations was 1\farcs3. 
A log of observations is given in Table 1, where $Z$ is the zenith angle and $Sky mag$ the extinction corrected sky
brightness (in mag~arcsec$^{-2}$). Reduction of the CCD data was 
performed in the standard manner using the ESO-MIDAS\footnote{MIDAS is 
developed and maintained by the European Southern Observatory.}
package. The $I$ band frames were obtained with exposure times
insufficient for detailed photometry, and we used these frames for
integral photometry only.

\begin{table}
\caption{ Characteristics of the optical observations. }
\begin{center}
\begin{tabular}{|c|c|c|c|c|}
\hline
 Dates & Band-        & Exp &  $Z$    & Sky  \\
      & pass         & (sec)& ($^{\circ}$)& mag. \\
\hline                   
13/14 Aug 2002 & $B$ & 3$\times$600 & 4  & 21.9\\
               & $V$ & 3$\times$600 & 14 & 21.3\\
               & $R$ & 3$\times$600 & 23 & 20.7\\
               & $I$ & 2$\times$600 & 6  & 20.1\\	       
\hline
\end{tabular}
\end{center}
\end{table}

\subsection{HI observations}

\begin{figure}
\centerline{\psfig{file=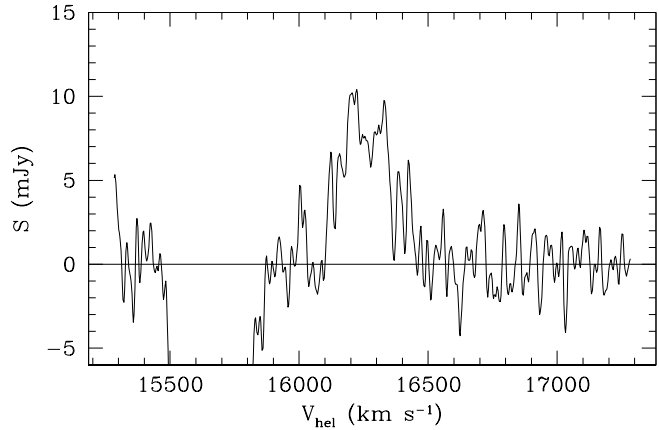,width=8.8cm,clip=}}
\caption{21 cm \HI\ line spectrum of ESO~474-G26. The negative dip in the 15~500 - 15~800 \kms\ range is due to imperfectly removed radio frequency interference.}
\end{figure}

The galaxy was observed in the 21 cm \HI\ line at the 100m-class 
Nan\c{c}ay\ decimetric radio telescope on 23, 25 and 28 August 2004, 
for a total of 1 hour per day. For further details on the telescope, 
data acquisition and data reduction methods,
including the off-line elimination of radio frequency interference (RFI),
see Monnier Ragaigne et al. (2001). The half-power beam width of the
telescope, of \am{3}{5}$\times$23$'$ ($\alpha$$\times$$\delta$), is
expected to cover the entire \HI\ distribution of the galaxy.

In each daily observation a sufficiently wide velocity range around the
galaxy profile was found to be free of RFI, in which the galaxy's \HI\ profile
was detected in both polarizations. Residual RFI causes the negative dip in the 15~500 - 15~800 \kms\ range in the averaged data (Fig.~2).

The average of the three spectra (see Fig.~2), smoothed to a velocity 
resolution of 18 \kms, has an rms noise level of 1.6 mJy. The galaxy 
profile has a peak flux density of 10 mJy, a center velocity 
\VHI=16~249$\pm$15 \kms, a velocity width at 50\% of peak maximum 
$W_{50}$=206$\pm$30 \kms, a velocity width
at 20\% of peak maximum $W_{20}$=254$\pm$46 \kms, and an integrated
line flux \IHI=1.8$\pm$0.2 \Jykms. These global \HI\ line parameters 
are directly measured values; no corrections have been applied to them 
for, e.g., instrumental resolution. We estimated the uncertainties 
in  \IHI, \VHI\ and the line widths following Schneider et al. 
(1986, 1990).

A comparison with the summed CO(1-0) spectra of Galletta et al. (1997) 
shows that the widths of the profiles are comparable.
The 470 \kms\ difference with the published central velocity of the
CO profile could be to be due to the application of the
relativistic Doppler shift formula to the CO data velocity, without the authors being aware of this correction -- converting to the conventional optical definition we use
($V=c$($\lambda$-$\lambda_0$)/$\lambda_0$) raises the CO center
velocity to 16~233$\pm$11 \kms, consistent with our \HI\ value and the
published optical radial velocities.

\section{Results}

\subsection{General structure}

ESO~474-G26 is a large system -- even the central galaxy (without the rings) is 
significantly larger and much more luminous than the Milky Way (Table~2).

The $V$-band image is displayed in Fig.~1 and represented as an isophotal plot in  Fig.~3. The overall optical morphology of ESO~474-G26 
is very interesting and intriguing: a nearly spherical central body
is surrounded by two almost perpendicular giant rings. Both ring-like
structures look somewhat irregular. 

The galaxy colors (after Galactic reddening correction
and $K$-correction for an early-type galaxy) correspond
to those of an Sb-type spiral. Fig.~4 presents the spectral energy distribution
(SED) for ESO~474-G26. As can be seen, although both observed and modeled SED curves for Sb 
galaxies fit well the observed SED from the $I$ band
to larger wavelengths, the SED for the prototype advanced merger remnant ``Atoms-for-Peace''
galaxy,NGC~7252 
(e.g., Schweizer 1982) 
gives a much better approximation (see further discussion). 

ESO~474-G26 is a strong source of far-infrared emission (Table~2).
The detected far-infrared luminosity, converted to a star formation
rate, indicates a very high rate of star formation (Table~2), as does
the large $L_{FIR}/L_B$ ratio of 1.1. The ratio of the H$_2$ mass to
blue luminosity (0.18~{\MsunLBsun) is typical of Sb--Sbc spirals 
(Young \& Knezek 1989), as are the relative \HI\ content and the H$_2$ to \HI\
ratios (Table~2).

A faint background galaxy (with a radial velocity of about 32000 km~s$^{-1}$ -- see 
the PRC) is located 80$''$ NW of ESO~474-G26 (Figs. 1,3). 
The total apparent magnitude of the galaxy is $V_T=17.1 \pm 0.1$ and its
colors are $(B-V)_T=+0.82$, $(V-R)_T=+0.30$, $(R-I)_T\approx0.0$.
Therefore, this is a giant (with an absolute magnitude in the $B$
band of about --20.7) late-type (from optical morphology and colors)
spiral galaxy. 

\begin{figure}
\centerline{\psfig{file=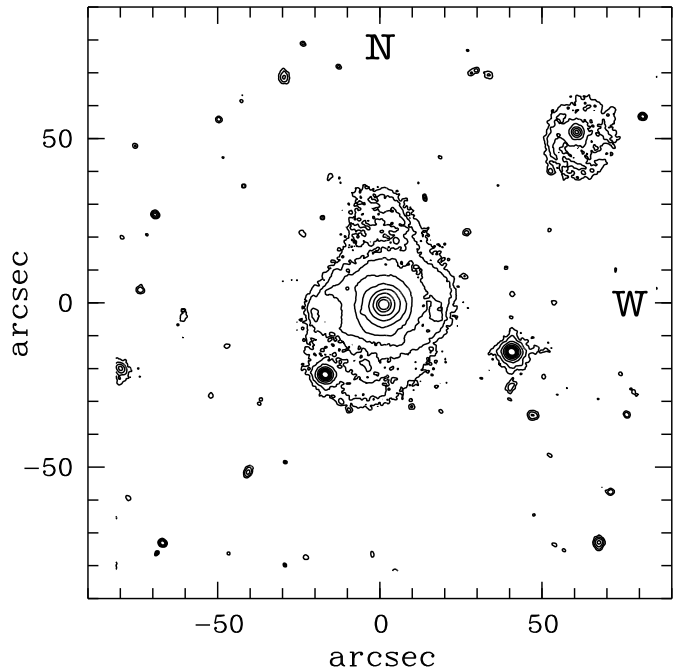,width=8.8cm,angle=0,clip=}}
\caption{Isophotal contour map of ESO~474-G26 in the $V$ band.
The faintest contour is 25.8 mag~arcsec$^{-2}$, and the isophote step --
0\fm75.}
\end{figure}

\begin{figure}
\centerline{\psfig{file=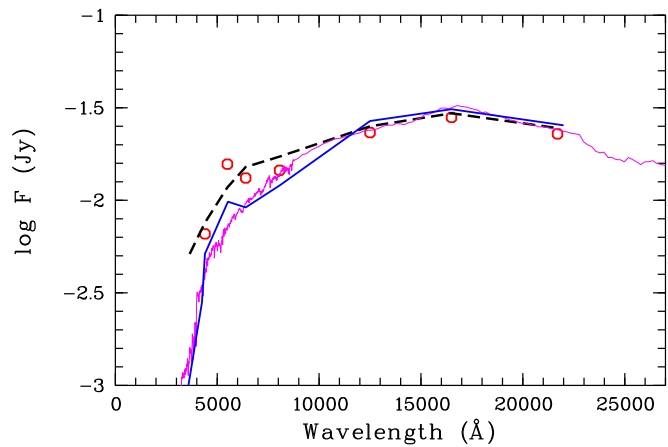,width=8.8cm,angle=-90,clip=}}
\caption{Spectral energy distribution for ESO~474-G26 (circles). The thick
solid line represents the distribution for the giant Sab/Sb galaxy
M~81 according to the NED data, the thin solid line the SED for an
Sb galaxy model (Fioc \& Rocca-Volmerange 1997), and the dashed line the SED
for the merger NGC~7252 (NED). All the SEDs have been  arbitrarily shifted to
fit approximately the observed SED of ESO~474-G26. $BVRI$ data are from our photometry.}
\end{figure}

General properties of ESO~474-G26 are summarized in Table~2.

\begin{table}
\caption{General properties of ESO~474-G26}
\begin{tabular}{lll}
\hline
Parameter &  Value & Ref. \\
\hline
Morphological type             & Sb: (merger)    &           \\
Heliocentric systemic velocity & 16~316 km~s$^{-1}$    & [1]   \\
Redshift, $z_{3K}$             & 0.05346         &  [2] \\
Adopted luminosity distance    &  238 Mpc        &       \\
($\Omega_m$=0.3, $\Omega_{\Lambda}$=0.7, $h$=0.7) & &\\
Scale                          & 1$''$=1.0 kpc & \\
Major axis, D$_{26}$~($\mu_B=26$) & 72$''$ (72 kpc) &               \\
Axial ratio, $b/a~(\mu_B=26)$   & 0.65           &              \\ \\
Total apparent                 &          & \\
magnitudes and colors:         &          & \\
$V_{\rm T}$                          & 13.98$\pm$0.06 &      \\
$(B-V)_{\rm T}$                      & +0.75$\pm$0.04 &   \\
$(V-R)_{\rm T}$                      & +0.49$\pm$0.04 &   \\
$(R-I)_{\rm T}$                      & +0.44$\pm$0.2 &   \\
$(J-H)_{\rm 2MASS}$            & +0.67          & [3]       \\
$(H-K)_{\rm 2MASS}$            & +0.35          & [3]       \\
Galactic absorption ($B$-band) & 0.085           & [4] \\
$K$-correction ($B$-band)      & 0.122           & [5] \\
Absolute magnitude, $M_B^0$    & --22.3         &     \\
                               &                &     \\ 
Central galaxy:                &                &     \\
Effective surface brightness, $\mu_e^0(B)$  & 21.5           &     \\
Effective radius, R$_e$ &      4.8 kpc & \\
Absolute magnitude, $M_B^0$    & --21.8         &     \\
V$_{\rm max}^0$ (km~s$^{-1}$)             &  175:          &     \\
Central galaxy - to - rings ratio &             &    \\
($B$-band)                        &  2:        & \\			       
                                  &             & \\
M(HI) (M$_{\odot}$)               & 2.2$\times$10$^{10}$ &     \\
M(H$_2$)   (M$_{\odot}$)           & 2.3$\times$10$^{10}$  & [6]\\	
$L_{FIR}$  ($L_{\odot}$)          & 1.6$\times$10$^{11}$ & [2],[7] \\
SFR$_{FIR}$ (M$_{\odot}$/yr)      & 43          &         \\
$L_{FIR}/L_B$                      & 1.1        &        \\
$L_{FIR}$/M(H$_2$) ($L_{\odot}$/M$_{\odot}$) & 7  & \\
M(H$_2$)/$L_B$ (M$_{\odot}$/$L_{\odot,B}$) & 0.18  & \\
M(HI)/$L_B$ (M$_{\odot}$/$L_{\odot,B}$) & 0.17  & \\
\hline \\
\end{tabular} 
[1] - HyperLEDA mean value;
[2] -- NASA/IPAC Extragalactic Database (NED);
[3] -- Skrutskie et al. (1997);
[4] -- Schlegel et al. (1998);
[5] -- Bicker et al. (2004);
[6] -- Galletta et al. (1997); 
[7] -- Based on the IRAS 60 and 100 $\mu$m  flux densities
\end{table}

\subsection{Central galaxy}

The central object has almost round isophotes (Fig.~3)
with apparent axial ratios in the $R$ passband of 
$\langle b/a \rangle$=0.94$\pm$0.015 within 15$''$ of the
center. Photometric profiles at two position angles are shown in
Fig.~5. At $\mid r \mid \leq 10''$ (or within 10 kpc
from the nucleus), the surface brightness can be well approximated
by the de Vaucouleurs' law. The observed colors of the galaxy
within $r=10''$ ($B-V=0.80$, $V-R=0.51$) are somewhat redder than
for the galaxy as a whole (Table~2) but bluer than 
typical values for ellipticals. 

\begin{figure*}
\centerline{\psfig{file=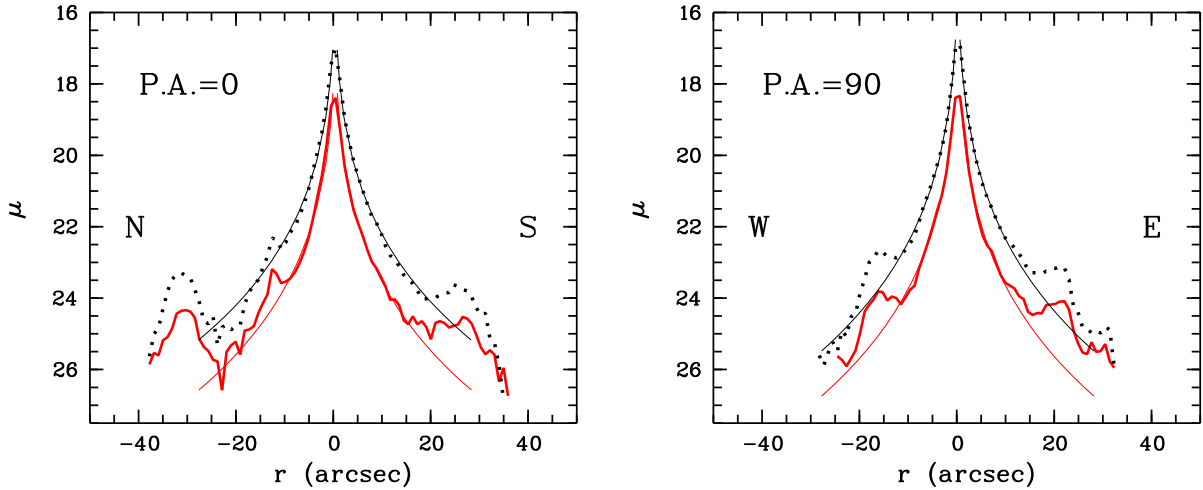,width=16cm,angle=-90,clip=}}
\caption{Photometric profiles for ESO~474-G26 at two position
angles. Thick solid lines represent the distributions in the
$B$ passband, and dotted ones those in the $R$. Thin lines show fits to  the 
profiles by the de Vaucouleurs' law.}
\end{figure*}

\begin{figure*}
\centerline{\psfig{file=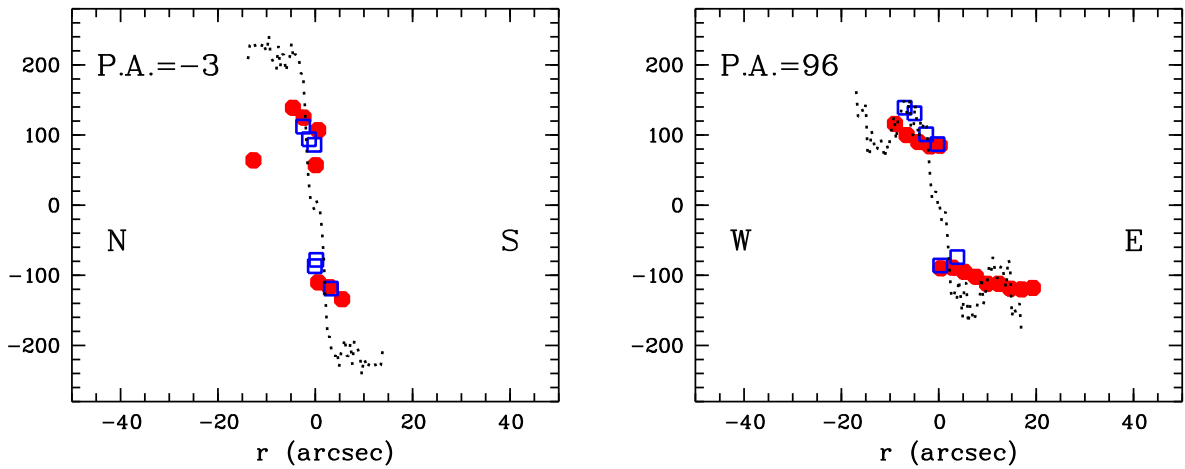,width=16cm,angle=-90,clip=}}
\caption{Emission-line radial velocities (in km~s$^{-1}$) along two position angles, according to the PRC. The filled circles are from H$\alpha$ measurements, and the squares are from [NII]6583~\AA\  measurements. Dotted lines represent the corresponding radial velocity distributions for the model galaxy (see Sect. 4). Observed velocities are from Whitmore et al. (1990), and are relative to heliocentric velocity of 16268 km~s$^{-1}$.}
\end{figure*}

A comparison of the effective parameters -- $\mu_e$, R$_e$ --
of the central galaxy with the Kormendy relation 
(photometric projection of the Fundamental Plane, Kormendy 1977) shows an enhancement in $\mu_e$ of about 1\fm0 -- 1\fm5 in $B$ at a fixed
R$_e$ (e.g., Capaccioli et al. 1992), or, at a fixed
surface brightness, that the galaxy has an effective radius 2--3 times larger
 than  ordinary ellipticals. 

Optical spectra showing the gas kinematics in ESO 474-G26 along three different position angles are given in the PRC. We reproduce in Fig.~6 the distributions of radial velocities at two position angles which are relatively close to the major axes of the rings. The central galaxy of ESO~474-G26 is surrounded by two ring-like structure, one polar and one equatorial. Both rings are in rotation around the galaxy center. Therefore, ESO~474-G26 can be formally classified as a kinematically confirmed PRG. The third spectrum published in the PRC shows the rotation of gas at a position angle of 132$^{\circ}$, which is intermediate between the two rings and reveals the complex kinematical structure of ESO~474-G26.

The observed maximum rotation velocity of the galaxy is 
V$_{max}\approx$140 km~s$^{-1}$ at $r\approx5''$ along P.A.=-3$^{\rm o}$ (PRC). 
The apparent axial ratio of this ring is $\langle b/a \rangle = 0.6 \pm 0.1$.
Assuming that the emission lines belong to the large ring, and that this ring is intrinsically circular, 
we can estimate its inclination as $i=53^{\circ} \pm 7^{\circ}$. Therefore, corrected for the inclination, the maximum rotation velocity at $r=5''=5$ kpc is V$_{max}^0\approx$175 km~s$^{-1}$.
The mass to luminosity ratio is  M/$L_B(r\leq 5'') \approx 1$~M$_{\odot}$/$L_{\odot,B}$, a value 
that is unusual for early-type galaxies and more typical of
spirals with active star formation. Due to the highly 
uncertain geometry of the galaxy, the above estimates of V$_{max}$ and
M/$L_B$ are tentative only.

\subsection{Rings}

The major axis position angle of the first (smaller) ring is 94$^{\circ}$, and its
diameter is $\approx37''$ or 37 kpc. The second (larger) ring has
P.A.=170$^{\circ}$ and diameter of 58$''$ (58 kpc). In projection,
the angle between the two rings is 76$^{\circ}$. 

The optical colors of the rings were determined through a 10$''$ diameter
circular aperture placed on the brightest parts of the rings along
its major axes. In the large ring we found the following colors at the indicated positions with respect to the nucleus:\\
$B-V=0.69$, $V-R=0.38$ (32$''$  N),\\ 
$B-V=0.56$, $V-R=0.36$ (27$''$  S). \\
and for the smaller ring: \\
$B-V=0.69$, $V-R=0.48$ (17$''$  W), \\
$B-V=0.54$, $V-R=0.43$ (21$''$  E). \\
Therefore, both rings are bluer than the central
galaxy (see previous section), and its extinction and redshift corrected
colors ($B-V\approx0.5$, $V-R\approx0.4$) are typical for Sc-Scd 
spirals (Buta et al. 1994; Buta \& Williams 1995) and similar to those of PRG
 rings (Reshetnikov et al. 1994, 1995).

Both rings show a color asymmetry, in the sense that their N and W parts
 are redder, which is not unusual for polar
rings -- see for instance 
the correlation found between the large-scale color asymmetry in the rings of the PRGs UGC~7576 and UGC~9796.

\begin{figure}
\centerline{\psfig{file=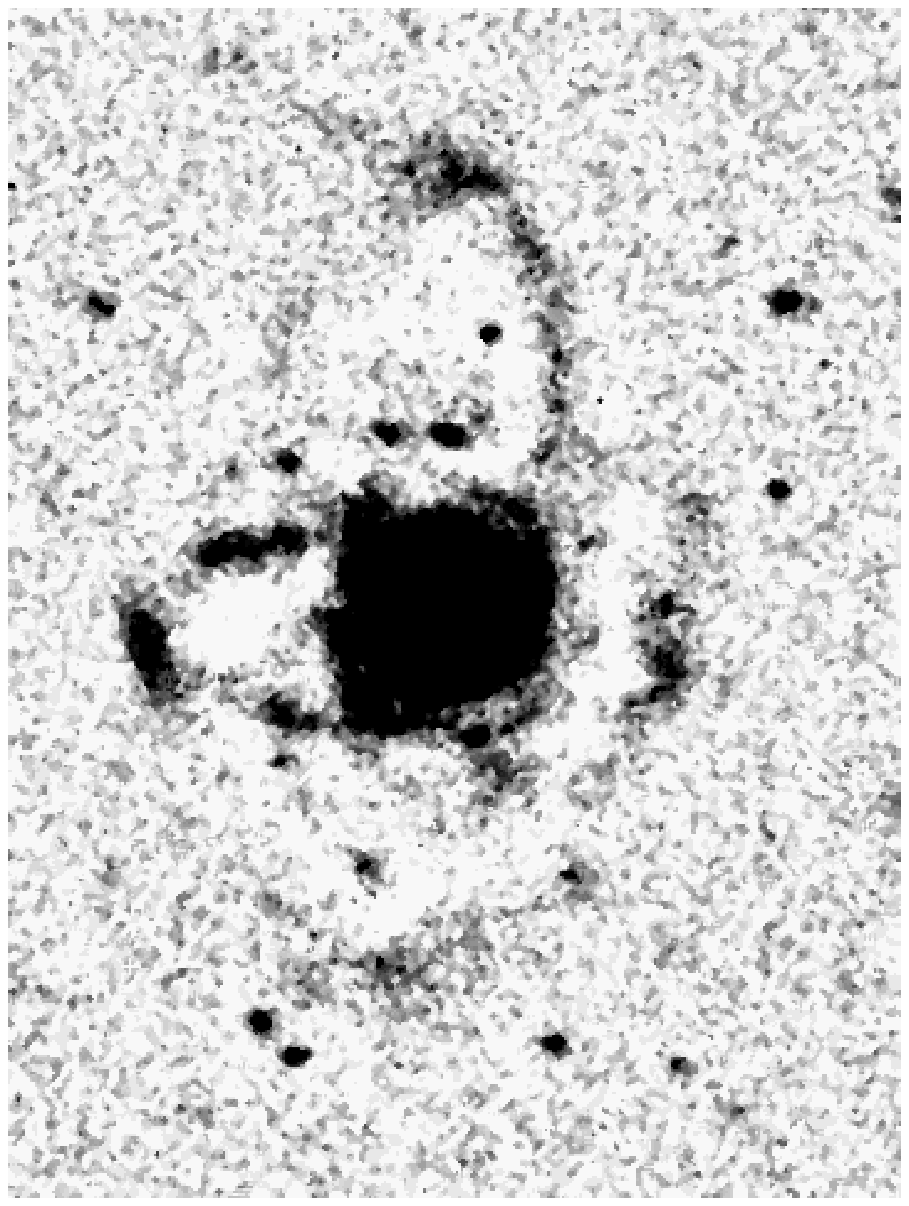,width=5.8cm,angle=0,clip=}}
\caption{Residual $R$ band image after the subtraction of the 8\farcs8$\times$8\farcs8
median filtered frame from the original image. The image size is  
$1\farcm44 \times 1\farcm08$. North is up, and East is left.}
\end{figure}

Various techniques of image enhancement have been employed to
reveal fine structure within ESO 474-G26 (see, for example,
Fa\'{u}ndez-Abans \& de Oliveira-Abans 1998, for a description of some
techniques). In Fig.~7 we present the residual image of the galaxy.
The smaller (east -- west) ring looks very irregular,
with many condensations. The brightest condensations show $B\approx21$
(or $M_B\approx-16$) and angular sizes around 2$''$ (2 kpc).   
Most probably, the condensations represent giant H\,{\sc ii}  regions.

The larger outer (north--south) ring has the southern part approaching (PRC, Galletta et al. 1997). The western
part of the ring is probably the nearest to us (from the
dust lane crossing the western side of the central body -- see Fig.~1).
In Fig.~8 we present a deep ($B+V+R$) and smoothed image of ESO~474-G26.
A faint ($\mu(V) \geq 26^m$) asymmetric envelope extends to
the east. The North and South edges of the large ring show extensions towards the east and west, respectively; they might be spiral arms.

\begin{figure}
\centerline{\psfig{file=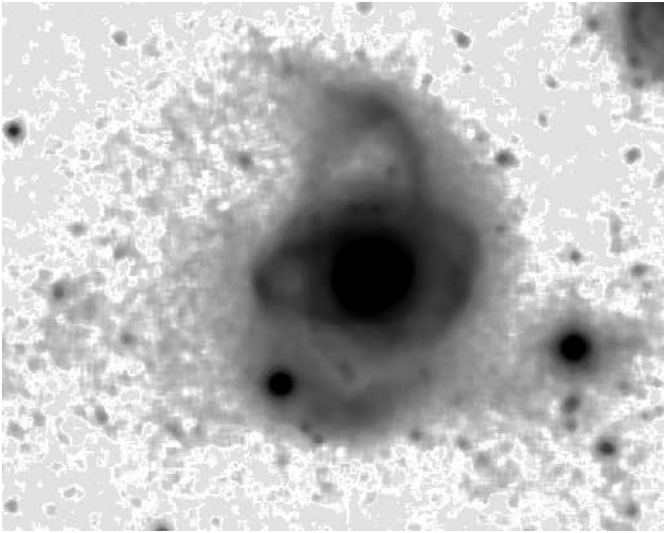,width=8.8cm,angle=0,clip=}}
\caption{Deep, high-contrast image of ESO~474-G26. The image size is  
$1\farcm8 \times 2\farcm1$.}
\end{figure}

\section{Discussion and conclusions}

The true geometrical structure and the possible origin of ESO~474-G26 
are intriguing mysteries. A number of formation scenarios -- resonance ring, accretion on a ringed early-type galaxy, and a minor merger -- can be excluded, for the following reasons:
\begin{itemize}
\item Although  the larger ring reveals some
spiral structure, it cannot be an outer pseudo-ring
of a resonant nature
(e.g., Buta \& Combes 1996), since  the central galaxy is clearly not the  barred disk galaxy required 
to form such a  ring. 
\item The irregular and unrelaxed aspect of the system could indicate a relatively recent external
polar accretion from a gas-rich donor galaxy onto a pre-existing
early-type galaxy with an outer ring. If so, one would  expect to find the relic of the donor
galaxy within a reasonable distance from ESO~474-G26, but 
ESO~474-G26 appears to be isolated and without a companion within 1 Mpc
projected distance at similar redshift.
\item A minor merger with a peculiar geometry
that formed the inner polar ring is unlikely to have resulted in the observed unperturbed outer
pseudo-ring. 
\end{itemize}

It is appropriate to test a major merger scenario that forms both rings at the same time, since, 
globally, ESO~474-G26 shows analogies to the prototypical
 merger remnant NGC~7252 (Schweizer 1982) 
-- which is also a giant galaxy, with an elliptical-like
surface brightness profile, relatively blue colors, peculiar kinematics,
 outer structures, and a similar spectral energy distribution (as shown in Fig.~4). 

To reproduce ESO~474-G26 in detail, we have to solve
a reverse problem: to find the characteristics and orientations
of the progenitor galaxies (as well as their orbits) from
the observed structure of the merger remnant. Of course, 
this  is not a fully determined problem, and many pieces of observational
data are missing, like the 2D gaseous and stellar kinematics of the system. Therefore,
we choose to find a generic solution only that  reproduces
a double ringed optical appearance with the right
relative proportions of the galaxy.

On the basis of the results by Bekki (1998) and BC03, we have considered a low-velocity head-on collision 
between two galaxies with orthogonal spiral disks. The simulations 
were performed with an $N$-body particle code which includes
gas dynamics, star formation and stellar mass-loss (see
Bournaud \& Combes 2002, 2003 for details). 
 
After several trials, we found an acceptable solution 
(Figs. 9 and 10). The main galaxy, hereafter denoted as the {\it victim} (as in Bekki 
1998), is a giant spiral galaxy (the exponential scalelength of the stellar disk is 
5.3 kpc) with a total mass M$_{\rm tot}=6.1 \times 10^{11}$~M$_{\odot}$, a
bulge-to-disk luminosity ratio 0.25, and a relatively low gas fraction 
M$_{\rm gas}$/M$_{\rm stars}$=0.03. The galaxy is surrounded by a moderately 
massive spherical dark halo with a mass ratio M$_{\rm DH}$/M$_{\rm stars}$=1.2. 
The second galaxy, the {\it intruder}, has a total mass of 
M$_{\rm tot}=2.7 \times 10^{11}$~M$_{\odot}$, an exponential scalelength of its 
stellar disk of 2.4 kpc,  M$_{\rm gas}$/M$_{\rm stars}$=0.065, and 
M$_{\rm DH}$/M$_{\rm stars}$=1.1. Initially, the galaxies were separated by 
70 kpc and their relative velocity was 70 km~s$^{-1}$. 

\begin{figure*}
\centerline{\psfig{file=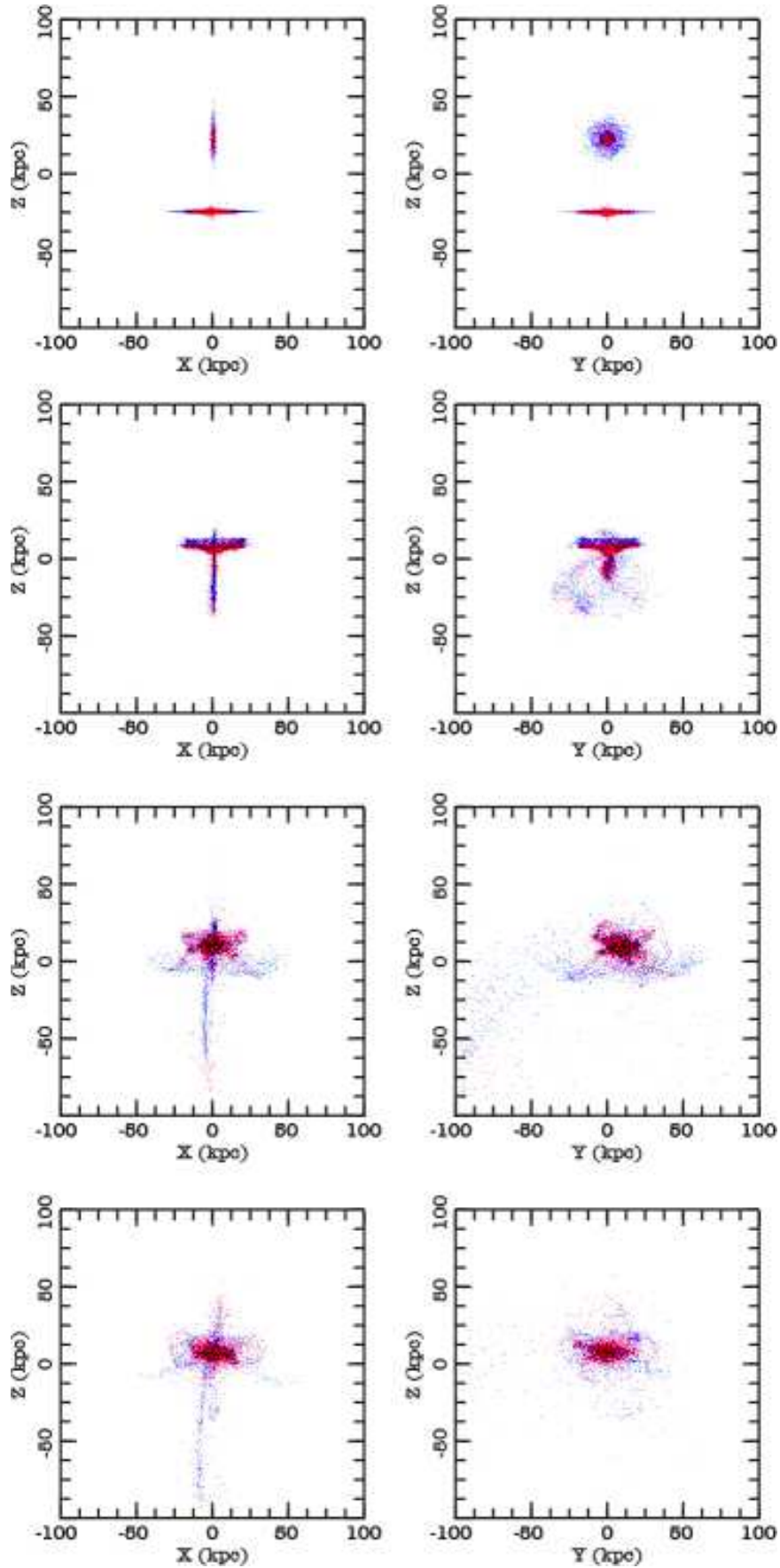,width=11.5cm,angle=270,clip=}}
\caption{Two orthogonal projections, (x,y) and (z,y), 
of the galaxy merging process  (one projection per  column).
Epochs after the beginning of the simulations are
$t$=200, 350, 500, and 650 Myrs (from top to bottom). Red points
represent stars, blue points represent gas.}
\end{figure*}

Fig.~9 shows the morphological evolution of the galaxies during the
merging process. The total number of particles used in this simulation  
is 2$\times$10$^5$. At $t$=700 Myr (or about 350 Myr
after the first crossing), we show in Fig.~10 a projection of the system
that is selected to mimic the observed double-ring appearance.
At this time, the merger remnant shows a ring-like structure 
along its apparent minor axis: this ring consists of tidal debris from 
the intruder, and is polar with respect to the victim disk. At the same 
time, an expanding collisional ring of the  Cartwheel type is formed by 
the victim galaxy material, which  is polar with respect to the intruder 
galaxy's initial plane. Thus, we find the two types of rings studied
in BC03 around the same object. The diameter of the collisional ring 
is 60 kpc, in agreement with the size of the large ring of 
ESO~474-G26 (Sect. 3.3). As one can see in Fig.~6, our model fits the observed 
rotation curve of ESO~474-G26 along the small ring, and is not inconsistent 
with the major axis kinematics, at least in the brightest regions of this 
ring. We show in Fig.~11 the composition of the two rings: the small, tidal 
ring is only made up of stars from the intruder galaxy (i.e., stars that 
initially were in the intruder disk, or stars formed recently from the intruder's 
gas), and the large ring is only made up of stars from the victim galaxy. 
The degree of mixing of the stellar populations in the rings is less than 2\%.

\begin{figure}
\centerline{\psfig{file=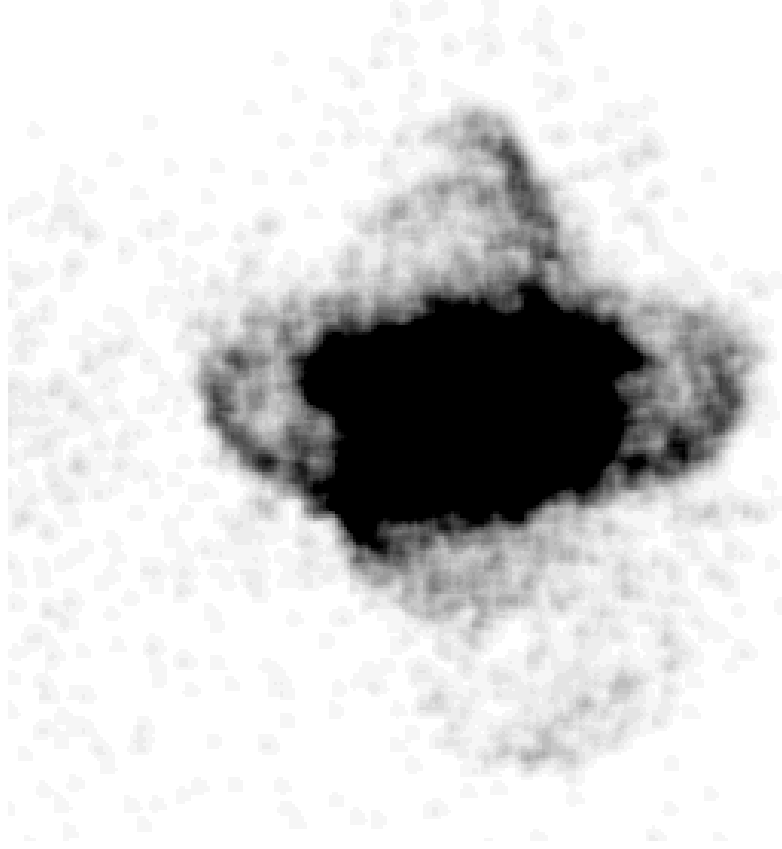,width=8.0cm,angle=0,clip=}}
\caption{Projected distribution of stars in the simulated merger 
at $t$=700 Myrs. The image size is 100 kpc $\times$ 100 kpc.
The projection was selected to fit the appearance of ESO~474-G26. }
\end{figure}

\begin{figure}
\centerline{\psfig{file=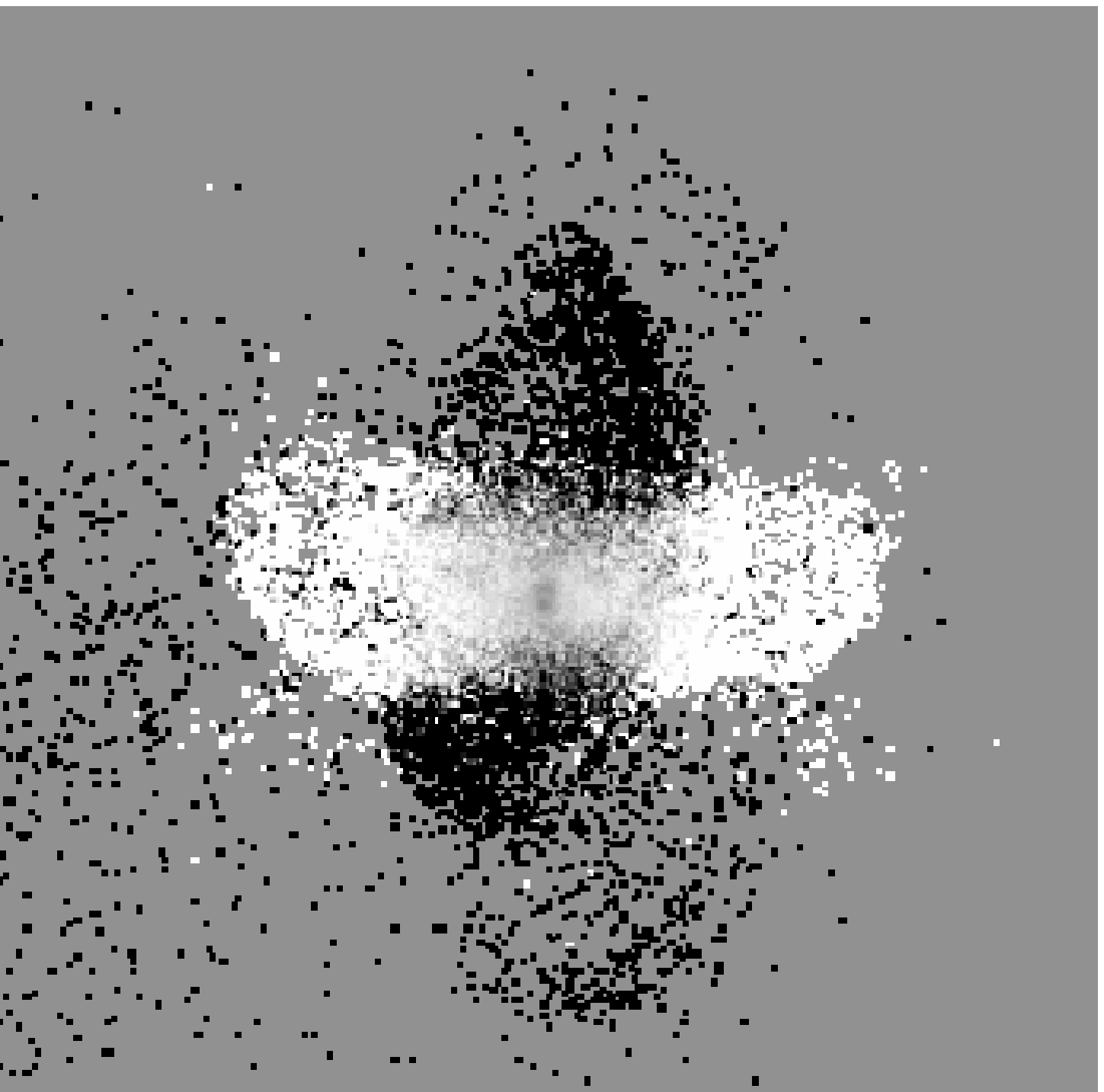,width=8.0cm,angle=0,clip=}}
\caption{Ratio map of the stellar populations from the victim and intruder 
galaxies. White: 100\% of the stars come from the victim disk (victim's stars 
or stars formed in gas from the victim disk) -- Black : 100\% of the stars 
come from the intruder's material -- Background Gray : 50\% from each galaxy, 
or no stars. The scale is linear. Each ring owes more than 98\% of its stellar 
mass to one of the pre-existing galaxies, and there is no mixing. In the central 500 pc, on the contrary, each galaxy contributes about 50\% to the stellar mass.}
\end{figure}

There are two differences between ESO~474-G26 and our artificial merger. 
The first  is that the ratio of the brightness of the rings to that of the peak in the central galaxy is half the value in the model to that actually observed. However, we have assumed a constant mass-to-light ratio in the model. As the rings mainly contain young stars, they could have a lower M/L ratio than the central galaxy, which solves this problem. The second difference is that the inner part of the model remnant looks unrelaxed in comparison to the observed smooth, elliptical-like surface brightness distribution in ESO~474-G26. 

In the simulations, the small ring, or the vertical coherent structure
(see Fig.~10) is a transient feature. It persists for about 200 Myr
and then disperses and vanishes. Approximately 1 Gyr after
the full merging, the remnant becomes smooth and elliptical-like.
Therefore, the impressive double-ring appearance of ESO~474-G26
could also be transient and not reflect a stable dynamical 
configuration.

It appears likely that we are indeed observing a transient stage in a particular merging scenario, since ESO~474-G26 is the only clear example with two such rings among $\approx 3 \times 10^4$ known galaxies with $B\leq$15\fm5 (PRC, NED).

We thus conclude that one of the most plausible solutions is that ESO~474-G26 represents a transient phase of a merging process between two spiral galaxies. This is based on its observational characteristics, which are typical of merger remnants, as well as on numerical simulations that  reproduce the general morphology of the galaxy.

\acknowledgements{We are grateful to the referee, C. Horellou, for her constructive comments. This work was carried out while VR was
visiting the LERMA in Paris, thanks to a CNRS 3-month visitor grant. 
VR acknowledges support from the Russian Foundation 
for Basic Research (03-02-17152) and from the Russian Federal Program 
``Astronomy'' (40.022.1.1.1101). 
M.F.-A. and M. de O.-A. acknowledge the partial support of the Funda\c{c}\~{a}o 
de Amparo \~{a} Pesquisa do Estado de Minas Gerais (FAPEMIG) and the 
Minist\'{e}rio da Ci\'{e}ncia e Tecnologia (MCT, Brazil).
The computations have been carried
out on the Fujitsu NEC-SX5 of the CNRS computing center, at IDRIS.
The Nan\c{c}ay Radio Observatory, which is the Unit\'e Scientifique de Nan\c{c}ay of
the Observatoire de Paris, is associated with the French Centre National
de Recherche Scientifique (CNRS) as USR B704, and acknowledges the
financial support of the R\'egion Centre as well as of the European Union.
}


\end{document}